\documentclass{article}
\usepackage{spconf,amsmath,epsfig}
\usepackage{amsbsy}
\usepackage{floatflt} 
\usepackage{amsmath}
\usepackage{amssymb}
\usepackage{times}
\usepackage{graphicx}
\usepackage{xspace}
\usepackage{paralist} 
\usepackage{setspace} 
\usepackage{xypic}
\xyoption{curve}
\usepackage{latexsym}
\usepackage{theorem}
\usepackage{ifthen}
\usepackage{subfigure}
\usepackage[ruled,vlined]{algorithm2e}
\usepackage{booktabs}
\usepackage{algorithmic}
\newtheorem{theorem}{Theorem}
\newtheorem{lemma}[theorem]{Lemma}
\title{Consensus in Correlated Random Topologies: Weights for Finite Time Horizon}
\twoauthors
  {Du$\breve{\mbox{s}}$an Jakoveti\'c, Jo\~ao Xavier\sthanks{ Work partially supported by the Carnegie Mellon$|$Portugal Program under a grant of the Funda\,c\~ao de Ci$\hat{\mbox{e}}$ncia e Tecnologia~(FCT) from Portugal. Du$\breve{\mbox{s}}$an Jakovetic holds a fellowship from~FCT.}}
	{Instituto Superior T\'ecnico~(IST)\\
	Instituto de Sistemas e Rob\'otica~(ISR)\\
	1049-001 Lisboa, Portugal}
  {Jos\'e M.~F.~Moura\sthanks{Work partially supported by  NSF under grant~\#~CNS-0428404, by  the Office of
Naval Research under MURI N000140710747, and by the Carnegie Mellon$|$Portugal Program under a grant of the Funda\,c\~ao de Ci$\hat{\mbox{e}}$ncia e Tecnologia~(FCT) from Portugal.}}
	{Carnegie Mellon University\\
	Department of Electrical and Computer Engineering\\
	Pittsburgh, PA 15213, USA}
\begin{document}
%
\maketitle
\begin{abstract}
We consider the weight design problem for the consensus algorithm under a finite time horizon.
We assume that the underlying network is random where the links fail at each iteration with certain probability
 and the link failures can be spatially correlated. We formulate a family of weight design criteria
 (objective functions) that minimize
 $n$, $n=1,...,N$ (out of $N$ possible) largest (slowest) eigenvalues of the matrix that describes
 the mean squared consensus error dynamics. We show that the objective functions are convex; hence,
  globally optimal weights (with respect to the design criteria) can be efficiently obtained.
   Numerical examples on large scale, sparse random networks
   with spatially correlated link failures show that: 1) weights obtained according to our criteria
   lead to significantly faster convergence than the choices available in the literature; 2)
different design criteria that corresponds to different $n$, exhibits very interesting tradeoffs: faster
transient performance leads to slower long time run performance and vice versa. Thus, $n$ is a valuable
degree of freedom and can be appropriately selected for the given time horizon.
\end{abstract}
\section{Introduction}
\label{section_introduction}
We consider the design of the weights for the consensus algorithm under a finite time horizon.
We assume the network
is random with links failing at each iteration with certain probability
 (see also~\cite{BoydFusion,JadbabaieErgodic,SoummyaChannelNoise}). The link failures are temporally
 uncorrelated
but can be spatially correlated, which is a better suited assumption for wireless
sensor networks (WSNs) than spatially uncorrelated
failures. 
Reference~\cite{BoydWeights} optimizes the weights for $static$ network topologies.
The weight design in~\cite{BoydWeights}
 leads to a convex problem of maximizing the algebraic connectivity of the (weighted) graph Laplacian with respect to
  the weights.
 In~\cite{journal}, we showed
  that the weight optimization for $random$ network topologies can
  be cast as a convex optimization problem. In this paper, we consider
  also the weight design for random topologies, but here consensus is over a finite number of iterations,
   i.e., under a $finite$ $time$ $horizon$. This problem is of interest in WSNs, where the number of iterations
    available can be limited to a small number due to the small power budget of sensors. Also, in certain applications, e.g., distributed detection of critical events (e.g., fire), result must be provided within certain critical time. Weight design with finite time horizon requires a new approach, different than~\cite{journal},
 since it must account for the transient phase of the consensus algorithm.
We first explain our methodology for solving the problem on static networks; we show that,
under a finite time horizon of $k$ iterations,
not only the slowest mode, but all the modes of the consensus error dynamics should be taken into account.
This leads to the formulation of a family of convex objective functions, indexed by $n$, that minimize the sum
of the $n$ largest eigenvaules that correspond to the $n$ slowest modes, $n=1,...,N$, of the error state matrix.
We generalize all
the results to random networks with spatially correlated links.
 We show that the objective functions are still convex for random topologies. Hence, globally optimal
weights (with respect to the defined criteria) can be efficiently obtained by numerical optimization.
The weight design~\cite{journal} is a special case of the functions family proposed here when $n=1$.
Numerical examples on sparse, large scale networks with spatially correlated link
failures show that: 1) the weights from our design family lead to significantly faster convergence than the available choices in the literature;
2) different choices from
our family (that correspond to different choice of $n$) exhibit very interesting tradeoffs: better transient performance leads to worse time asymptotic performance and vice versa.
Thus, depending on the given time horizon, one can choose the appropriate cost function (i.e., $n$) to achieve the desired performance.
\section{Problem model}
We follow the model of a random network as in~\cite{SoummyaTopologyDesign}, except that we assume that the links
 can be spatially correlated, while in~\cite{SoummyaTopologyDesign} they are uncorrelated.
  We briefly introduce relevant objects and
 notation. The supergraph $G = (V,E)$
  is the graph that collects all the links
 with non zero probability of being alive. ($V$ is the set of nodes, and $E$ is the set of
  undirected links.)
At any time step $k$: 1) link $\{i,j\} \in E$ is active with probability $P_{ij}$;
2) link $r$, incident to nodes $i$ and $j$, and link $s$, incident to nodes $l$ and $m$,
are correlated with the corresponding cross variance $R_{rs}$.
Consensus is an iterative distributed algorithm that computes the average
$x_{\mbox{\scriptsize{avg}}} = \frac{1}{N} \sum_{i=1}^N x_i(0)$
of scalar sensor measurements (or some other data) $x_i(0)$ iteratively at each sensor $i$:
\begin{equation}
\label{eqn_consensus}
x_i(k+1)= \left( 1- \sum_{j \in {O}_i} \mathcal{W}_{ij}(k) \right) x_i(k)+ \sum_{j \in {O}_i} \mathcal{W}_{ij}(k) x_j(k)
\end{equation}
In~\eqref{eqn_consensus}, $O_i$ denotes the neighborhood set of sensor $i$, i.e.,
$
{O}_i = \{  j:\,\, \{ i,j \} \in E \}.
$
Defining the state matrix $\mathcal{W}(k)=\left[ \mathcal{W}_{ij}(k) \right]$ and the state vector $x(k)=(x_1(k),...,x_N(k))^T$ we have in compact form:
$
x(k+1) = \mathcal{W}(k)\,x(k).
$
Also, it is straightforward to show (e.g.,~\cite{BoydWeights}) that the consensus
error $e(k)=x(k)-x_{\mbox{\scriptsize{avg}}} \,1$ follows the dynamics:
\[
e(k+1) = \left( \mathcal{W}(k)  -  J \right) e(k),\,\,J=\frac{1}{N}11^T.
\]
We consider the case when $\mathcal{W}_{ij}(k)$, $\{i,j\} \in E$, is equal to a prescribed number $W_{ij}$ whenever link $\{i,j\}$ is alive
 and zero otherwise. Thus, $\mathcal{W}_{ij}(k)$ is a binary random variable for $\{i,j\} \in E$ ($\mathcal{W}_{ij}(k)=0$ if $\{i,j\} \notin E$).
We design the weights $\{W_{ij}\} = \{W_{ij} \in \mathbb{R}:\,\, \{i,j\} \in E, \,\,i <j \}$ that lead to fast average consensus under a finite time horizon. We define a family of convex objective functions (criteria) that lead to fast
consensus in random topologies. We first explain our methodology in the context of a static topology (section~\ref{section_motivation}) and then consider correlated random topologies (section~\ref{section_random_topology}).
%
%
%
\section{Static topology: Motivation example}
\label{section_motivation}
We consider first that the network is static.
Then, the state matrix $\mathcal{W}(k)=\mathcal{W}$ is deterministic and is given by: $\mathcal{W}_{ij}=W_{ij}$, $\{i,j\} \in E$; $\mathcal{W}_{ij}=0$, $\{i,j\} \notin E$; $\mathcal{W}_{ii}=1-\sum_{i \in O_i}W_{ij}$.
We compute the eigenvalue decomposition of the matrix $\mathcal{W}-J$,
$\mathcal{W}-J = Q\,\Lambda\,Q^T,$
where the eigenvalues are ordered such that $|\lambda_1| \geq |\lambda_2| \geq... \geq |\lambda_N| = 0$ ($\lambda_N=0$ since $\mathcal{W}1=1$ and $J1=1$.)
A necessary and sufficient condition for the consensus algorithm~\eqref{eqn_consensus} to converge is that $|\lambda_1|<1$~\cite{BoydWeights}.
The consensus error can be written as:
\begin{equation}
\label{eqn_error_modes}
\|e(k)\|^2 = \sum_{i=1}^{N-1} \lambda_i^{2k}\,\left( q_i^T e(0)\right)^2 = \sum_{i=1}^{N-1} \zeta_i^2(k)
\end{equation}

Weight optimization for static topology has been studied in~\cite{BoydWeights}. This reference proposes two different criteria (objective functions) to optimize the weights, the time asymptotic convergence rate $r_{\mathrm{as}}$ and the worst case per step convergence rate $r_{\mathrm{step}}$ defined as:
\begin{eqnarray*}
r_{\mathrm{as}} &=& \sup_{e(0)\neq 0} \lim_{ k \rightarrow \infty} \left(  {\|e(k)\|}/{\|e(0)\|} \right)^{1/k} \\
r_{\mathrm{step}} &=& \sup_{e(k) \neq 0}  {\|e(k + 1)\|}/{\|e(k)\|}
\end{eqnarray*}
Since the matrix $\mathcal{W}-J$ is symmetric, we have that $r_{\mathrm{as}} = r_{\mathrm{step}} = |\lambda_1|$,~\cite{BoydWeights}. Thus, $r_{\mathrm{as}}$ and $r_{\mathrm{step}}$ both map to
the minimization of $|\lambda_1|$ with respect to the weights $\{W_{ij}\}$.
This is a convex optimization problem~\cite{BoydWeights}.


We argue that for small $k$ and for the optimal average performance, rather than the worst case performance,
a criterion for minimization different than $| \lambda_1 |$ should be considered. We give a motivational numerical example by considering a (static) connected network with $N=120$ nodes and $M=449$ edges. Figure 1 plots $\|e(k)\|^2$
averaged over 1000 different random initial conditions for two different weight choices: 1) the weights that minimize $|\lambda_1|$; 2) the Metropolis weights (MW),~\cite{BoydFusion}. Metropolis weights are a heuristic weight choice and thus not optimal.  However, in first 20 iterations, MW performs better.
The reason is that minimization of $|\lambda_1|$ causes several other eigenvalues of $\mathcal{W}-J$ to be close in modulus to $\lambda_1$. Eqn.~\eqref{eqn_error_modes} clearly shows that, for a small number of iterations $k$, all nonzero eigenvalues $\lambda_i$ affect the error (since for small $k$ $\lambda_i^{2k}$ are not negligible, $i=1,...,N-1$ ). Thus, for small $k$, it is better to have many eigenvalues of $\mathcal{W}-J$ small in modulus than to minimize $|\lambda_1|$ at the cost of having large $\lambda_2, \lambda_3,...$.
\begin{figure}[thpb]
      \centering
      \includegraphics[scale=0.23 ]{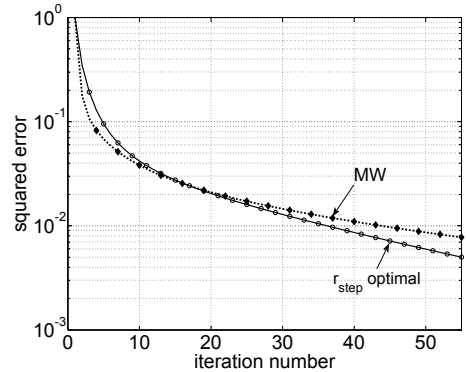}
      \caption{   Squared error versus iteration $k$ for static network.      }
      \label{figure_1}
\end{figure}
In order to make all modes $\zeta_i(k)$ (eqn.~\eqref{eqn_error_modes}) small, we propose to minimize
the sum of the squares of the eigenvalues $\lambda_i$, i.e., to minimize the function
 $\psi_{N}\left(  \{ W_{ij} \}  \right) := \sum_{i=1}^{N-1} \lambda_i^2 = \mathrm{tr} \left(  (W-J)^2\right)$.
Further, we may reason as follows. For $k$ being very
large, only the largest eigenvalue is of interest; for $k$ being
very small, all the eigenvalues should be taken into account.
For some medium range of the number of iterations, it is reasonable
to try to minimize the $n$ largest eigenvalues of
$\left(\mathcal{W} - J\right)^2$, $1 < n < N$. This leads to the
minimization of function $\psi_n \left( \{W_{ij}\} \right) : = \sum_{i=1}^n \lambda_i^2$, i.e., to
the following
optimization problem:
\begin{equation}
\begin{array}[+]{ll}
\mbox{minimize} & \psi_n \left( \{W_{ij}\} \right) \\
\mbox{subject to} & W_{ij} \in {\mathbb R}, \,\, \{i,j\} \in {E}\\                & | \lambda_1 | < 1
\end{array}
\label{eqn_optimization_k_largest}
\end{equation}
The constraint $|\lambda_1| <1 $ assures that we search only over the weight choices for which the
consensus algorithm converges. It can be shown (the proof is omitted here) that the functions $\psi_n(\cdot)$, $n=1,...,N-1$, are convex,
and thus~\eqref{eqn_optimization_k_largest} is a convex problem. 
%
%
\begin{lemma}
The function $\psi_n \left(  \{W_{ij} \} \right)$
is convex for
any $n=1,...,N-1$.
\label{Lemma_1}
\end{lemma}
\section{correlated random topology}
\label{section_random_topology}
We generalize the results from the previous section to the case
of random network topology with spatially correlated link failures.
Reference~\cite{journal} studies the weight design for correlated random topology. Denote the consensus error covariance matrix by $\Sigma(k) = \mathrm{E} \left[ e(k)e^T(k)\right] $. It can be shown that~\cite{journal}:
\begin{equation}
\mathrm{tr} \left(  \Sigma(k+1) \right) = \mathrm{tr}  \left(  \Sigma(k) \,\left( \mathrm{E}\left[ \mathcal{W}^2\right] -J\right) \right)
\end{equation}
Reference~\cite{journal} minimizes $\phi_1 \left( \{ W_{ij} \} \right) = \lambda_1 \left( \mathrm{E}\left[ \mathcal{W}^2 \right] - J \right)$. This quantity represents:
1) the worst case per step mean squared rate of convergence (eqn.~\eqref{eqn_WC});
2) the upper bound on the time asymptotic convergence rate (eqn.~\eqref{eqn_TA}), see~\cite{whichShouldI}:
\begin{eqnarray}
\label{eqn_WC}
\sup_{ \mathrm{E} \left[  e(k)e(k)^T\right] \succeq 0,\, \mathrm{E} \left[  e(k)^Te(k) \right] \neq 0 }
\frac{\mathrm{E} \left[  e(k+1)^Te(k+1) \right] }{   \mathrm{E} \left[  e(k)^Te(k) \right]    } \\
\label{eqn_TA}
\lim_{k \rightarrow \infty}
\frac{1}{k} \mathrm{ln} \left(  \frac{\| e(k) \|} {\|e(0)\|}  \right)^{1/k}
\leq 0.5\, \mathrm{ln} \left(     \lambda_1 \left(    \mathrm{E} \left[  \mathcal{W}^2 \right]-J    \right) \right)
\end{eqnarray}
%
Define the function
\begin{equation}
\phi_n \left( \{W_{ij} \} \right) = \sum_{i=1}^n \lambda_i\left( \mathrm{E} \left[ \mathcal{W} ^2\right]  -  J  \right).
\end{equation}
We remark that $\phi_1(\cdot)$ for random topology boils down to $\psi_1(\cdot)$ for static topology.
Thus, minimization of $\phi_1$ boils down to minimization of $|\lambda_1(\mathcal{W}-J)|$ if the network is static.
The same holds for the functions $\phi_n(\cdot)$ and $\psi_n(\cdot)$, $n=2,...,N-1$. This is because the matrix $\mathrm{E} \left[ \mathcal{W}^2 \right]-J$ is simply the matrix $\mathcal{W}^2 - J = (\mathcal{W}-J)^2$ when the network is static.
Thus, we propose
to solve the following optimization problem:
\begin{equation}
\begin{array}[+]{ll}
\mbox{minimize} & \phi_n\left( \{ W_{ij} \} \right)\\
\mbox{subject to} & W_{ij} \in {\mathbb R}, \,\, \{i,j\} \in {E}\\
                  & \phi_1\left(  \{W_{ij}\} \right) < 1
\end{array} \label{eqn_optimization_problem_random_k_largest}
\end{equation}
Constraint $\phi_1\left( \{W_{ij}\}\right)<1$ restricts the search only over the points $\{W_{ij}\}$
for which the algorithm converges in mean squared sense.
Special case $n=1$ is studied in~\cite{journal}. We have the following result:
\begin{lemma}
\label{lema_druga}
The function $\phi_n\left( \{W_{ij}\}\right),\,n=1,...,N-1$, is convex.
\end{lemma}
The proof of Lemma 2 for $n=1$ is in~\cite{journal}, but we extend it to the case of arbitrary $n$, $n=2,...,N-1$. Due to lack of space it is omitted. In view of Lemma 2, optimization problem~\eqref{eqn_optimization_problem_random_k_largest} is convex; hence, globally optimal
 $\{W_{ij}\}$ can be efficiently obtained.
%
%
\section{Simulations}
We consider a sparse geometric supergraph with $N=120$ nodes and $M=449$ edges. Nodes are uniformly distributed on a unit square and the pairs of nodes with distance smaller than a radius $r$ are connected by an edge. We define the formation probabilities by the following model: $
P_{ij} = 1-c_1\, \left( \delta_{ij}/r \right)^2$, $\{i,j\} \in E,\,\,c_1=0.6.
$.
 Link $r$, incident to nodes $i$ and $j$, and link $s$, incident to nodes $l$ and $m$ (and $P_{ij}<P_{lm}$) are correlated at time $k$; the corresponding cross-variance is given by $R_{rs} = c_2\,P_{ij} \left(  1 - P_{rs} \right)$, $c_2=0.2$. The correlated binary random links are simulated by the method in~\cite{Quadish}.
We compare the performance of our solutions with the weight choices for random topologies previously proposed in the literature, namely with the Metropolis weights~\cite{BoydFusion}, and the weights proposed in~\cite{whichShouldI}, which we refer to as the supergraph based weights (SGBW).
Figure 2 plots the mean squared error averaged over 100 different initial conditions. We compare the following weight choices:
1) MW; 2) SGBW; 3) weights obtained by minimizing
$\phi_1$ (which also appear in~\cite{journal}); 4) weights obtained by minimizing $\phi_{30}$. Numerical minimization
 of~\eqref{eqn_optimization_problem_random_k_largest} is done
 by the subgradient algorithm for constrained minimization: if the current point $\{W_{ij}\}$ is feasible
 ($ \phi_1 \left( \{ W_{ij}\}\right) < 1$), we compute the subgradient step in the direction of the objective function $\phi_n$;
 2) if the current point $\{W_{ij}\}$
 is infeasible ($ \phi_1 \left( \{ W_{ij}\}\right) \geq 1$), we compute the subgradient step in the direction of $\phi_1$ (constraint function).
\begin{figure}[htb]
\begin{minipage}[b]{1.0\linewidth}
  \centering
  \includegraphics[scale=0.25]{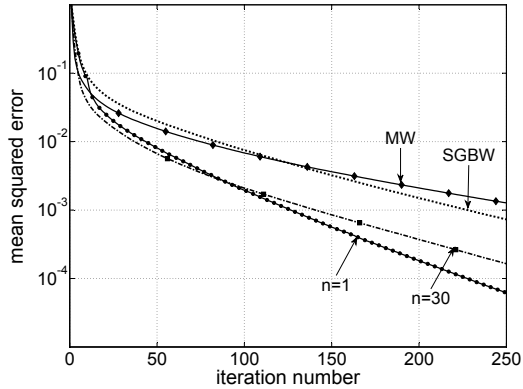}
  \centerline{(a) Comparison of $\phi_{30}$ and $\phi_1$ with MW and SGBW}\medskip
\end{minipage}
\begin{minipage}[b]{\linewidth}
  \centering
 \includegraphics[scale=0.25]{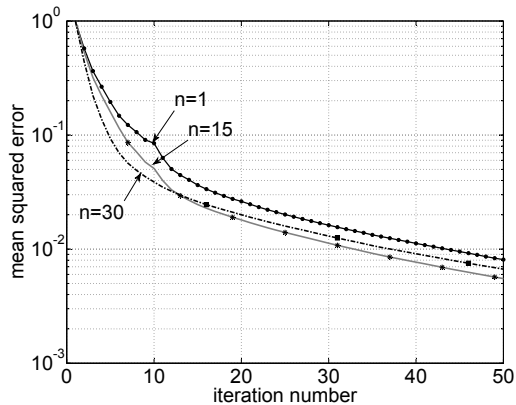}
  \centerline{(b) Tradeoff in choice of $\phi_1$, $\phi_{15}$, $\phi_{30}$}\medskip
\end{minipage}
\caption{Random network: mean squared error versus iteration number averaged over 100 random initial conditions.}
\label{fig:res}
\end{figure}
Figure 2 (a) shows that both $\phi_1$ and $\phi_{30}$ outperform SGBW and MW. To decrease the error to $1\%$, $\phi_1$ takes around 44 iterations; $\phi_{30}$
takes 37 iterations; SGBW and MW take more than 75 iterations to achieve $1\%$ precision.
We see that $\phi_1$ and $\phi_{30}$ exhibit
a tradeoff: in the transient regime (i.e., for small iterations $k$), $\phi_{30}$ performs better;
 for large $k$, $\phi_1$ performs better.
For the precision of $1\%$, $\phi_{30}$ is a better choice (it saves 7 iterations compared to $\phi_{1}$, see also Figure 2(b));
 for the precision of $0.1\%$, $\phi_1$ is a better choice (it saves around 15 iterations compared to $\phi_{30}$) (see Figure 2(a)).
Figure 2(b) presents the performance for 3 different choices of $n$, $n=1,\,n=15,\,n=30$, in initial 50 iterations. We see that, for the $1\%$ precision, $\phi_{15}$ reduces by $25\%$ the number of iterations compared to $\phi_1$, from  $43$ to $33$.
Possibility of choosing different $n$ is valuable in practice.
One can envision the application of the family $\{\phi_n\}$, for instance, in
tracking applications, where combined technique of detection and estimation is used.
In the first phase of tracking, target should
 be detected roughly in an area. This task can be done by distributed detection using consensus algorithm~\cite{moura_detection}.
 For this task, by nature of problem, high precision is not required, and thus one should choose $\phi_{30}$
 criterion for fast solution.
 In the second phase of tracking, target trajectory is estimated, which can be done distributively
 based on consensus algorithm~\cite{BoydFusion}. This task requires higher precision. For this phase,
 one could choose $\phi_{1}$ or $\phi_{15}$.
%
%
\section{CONCLUSION}
In this paper, we studied the weight design for a finite time horizon
consensus with
random topology and spatially correlated link failures.
We addressed the problem of finding the
optimal weights that yield the best average
performance of the algorithm. We consider a
 finite time horizon, i.e.,
only a limited number of consensus iterations is
available. We formulate a class of optimization problems for
weight design under a finite time horizon.
This class minimizes the
sum of the $n$ largest eigenvalues of the matrix that describes the mean
 squared error dynamics
 , $n=1,...,N$.
We show that the optimization
problem is convex for arbitrary $n$ and hence can be efficiently globally solved.
 Numerical examples on large scale, sparse graphs with spatially
 correlated link failures show that,
 for any choice of $n$,
optimization provides solutions better than the weight choices
previously proposed in the
literature.
Also, the weight optimization for finite time consensus leads to
very interesting tradeoffs: larger
$n$ yields faster convergence in the transient regime and slower
convergence in the long run regime. The parameter $n$ represents a valuable
degree of freedom than can be appropriately set for given time horizon.

\bibliographystyle{IEEEtran}
\bibliography{IEEEabrv,Bibliography}
\end{document}